# Data-Driven Catalyst Design: A Machine Learning Approach to Predicting Electrocatalytic Performance in Hydrogen Evolution and Oxygen Evolution Reactions

Vipin K E[1] and Prahallad Padhan[1,2]

[1]*Department of Physics, Nanoscale Physics Laboratory, Indian Institute of Technology Madras, Chennai 600036, India*

[2]*Functional Oxides Research Group, Indian Institute of Technology Madras, Chennai 600036, India*

## Abstract

The transition to sustainable green hydrogen production demands innovative electrocatalyst design strategies that can overcome current technological limitations. This study introduces a comprehensive data-driven approach to predicting and understanding catalytic performance for Hydrogen Evolution Reaction (HER) and Oxygen Evolution Reaction (OER) using advanced machine learning methodologies. By usimg a dataset of 16,226 data points from the Catalysis-hub database, we developed a novel stacking ensemble model that integrates Random Forest, XGBoost, and Support Vector Regression to predict Gibbs free energy of adsorption across diverse bimetallic alloy surfaces. Our innovative feature engineering strategy combined Matminer-based compositional analysis, Principal Component Analysis for adsorption site related features, and correlation screening to generate robust predictive descriptors. The machine learning model demonstrated exceptional predictive capabilities, achieving $R^2$ values of 0.98 for HER and 0.94 for OER, with Mean Absolute Error values of 0.251 and 0.121, respectively. Shapley Additive Explanations (SHAP) analysis revealed critical insights into the complex interplay of compositional, structural, and electronic features governing catalytic performance. The research provides a powerful computational framework for accelerating electrocatalyst design, offering unprecedented insights into the fundamental properties that drive hydrogen evolution and oxygen evolution reactions. By bridging advanced machine learning techniques with fundamental electrochemical principles, this study presents a transformative approach to developing cost-effective, high-performance catalysts for sustainable hydrogen production.

## Introduction

The global energy landscape stands at a critical juncture, with an urgent need to transition from fossil fuel-dependent systems to sustainable, clean energy solutions. Green hydrogen emerges as a important technology in this transformation, offering a versatile and zero-carbon energy carrier that can potentially decarbonize hard-to-abate sectors such as heavy industry, long-distance transportation, and energy-intensive manufacturing [1]. At the heart of this technological revolution lies Proton Exchange Membrane (PEM) electrolysis, a sophisticated water-splitting technique that converts renewable electricity into high-purity hydrogen with remarkable efficiency. PEM electrolysis represents a technological marvel of electrochemical engineering, capable of operating at high current densities and demonstrating exceptional responsiveness to intermittent renewable energy sources like solar and wind [1,2]. Unlike traditional hydrogen production methods that rely on fossil fuel-based processes, PEM electrolysis offers a genuinely sustainable pathway. The technology's ability to produce extremely pure hydrogen, coupled with its potential for compact system design and high-pressure operation, makes it particularly attractive for various applications, ranging from transportation fuel cells to industrial processes.

The core challenge in PEM electrolysis lies in the fundamental electrochemical reactions: the Hydrogen Evolution Reaction (HER) at the cathode and the Oxygen Evolution Reaction (OER) at the anode. These reactions, while seemingly straightforward, involve complex electron transfer mechanisms that significantly impact the overall efficiency and economic viability of hydrogen production [3]. The HER, a two-electron process ($2H^+ + 2e^- \rightarrow H_2$), occurs relatively quickly and with lower overpotential compared to the OER. The OER, on the other hand, is a more complex four-electron process ($H_2O \rightarrow 1/2 O_2 + 2H^+ + 2e^-$) and is often considered the limiting factor in water splitting due to its slower kinetics and higher overpotential.

Current catalyst technologies predominantly rely on precious metals, which pose significant economic and scalability challenges. Platinum remains the benchmark catalyst for HER due to its optimal hydrogen binding characteristics, while iridium and ruthenium oxides dominate OER catalysis in acidic environments [4,5]. However, the astronomical costs and extreme scarcity of these metals create insurmountable barriers to large-scale green hydrogen implementation. A single gram of iridium, for instance, can cost over $5,000, making widespread deployment economically unfeasible.

The scientific community has responded by exploring earth-abundant alternatives, investigating materials like nickel-based alloys, molybdenum sulfides, and cobalt phosphides as potential platinum substitutes [4]. These efforts have yielded promising initial results, but significant challenges remain in achieving comparable catalytic performance, particularly in maintaining stability under the harsh acidic conditions inherent to PEM electrolysis.

In the rapidly evolving landscape of materials science, data-driven machine learning (ML) has emerged as a transformative approach to material design across diverse research domains. Over the past decade, computational techniques have revolutionized exploration in critical fields such as superconducting materials, thermoelectric technologies, and electrochemical catalysis [8-13]. The machine learning approach offers multifaceted advantages in catalyst research. Computational models can rapidly screen thousands of potential material combinations, identifying non-intuitive structures and compositions that might be overlooked through conventional research methods [7]. Deep learning algorithms can be trained on existing experimental datasets, developing predictive capabilities that extract complex structure-property relationships beyond human intuition.

The recent proliferation of ML-aided methodologies has significantly expanded our ability to explore potential electrocatalysts, enabling unprecedented insights into the intricate chemical and physical properties of material systems [14-16]. Researchers have developed sophisticated computational frameworks that push the boundaries of traditional materials discovery. Notable approaches include Chen et al.'s universal ML framework for screening HER catalysts using graph convolutional neural networks, which successfully analyzed 43 potential catalyst candidates [17]. Pandit et al. pioneered a hybrid approach, integrating density functional theory (DFT) with supervised machine learning to uncover novel heterogeneous NiCoCu-based HER catalysts [18]. Similarly, Wu et al. implemented advanced deep learning techniques with crystal graph convolutional neural networks to expedite the identification of high-performance two-dimensional HER catalysts from extensive material databases [19].

A critical challenge in these computational approaches lies in constructing robust descriptors that can effectively establish meaningful relationships between material features and catalytic properties [20,21]. Researchers have developed innovative strategies to address this complexity. Wang et al. introduced a sophisticated surface center-environment feature model that combined DFT and ML methods to investigate perovskite performance in oxygen evolution reactions [22]. Sun et al. conducted a comprehensive investigation of the HER process on graphdyine-based atomic catalysts, performing intricate feature analysis of adsorption energies, electronic structures, reaction pathways, and active sites [23].

The advent of ab initio calculation methods, particularly density functional theory, has revolutionized the virtual design of catalysts through advanced computer simulations [24]. While DFT and Monte Carlo simulations offer unprecedented capabilities in determining catalytic reaction kinetics, including activation states and energy contributions [25,26], researchers recognize the inherent limitations of comprehensive first-principles evaluations.

To navigate these challenges, the scientific community has developed strategic approaches based on rational descriptor identification. The Brønsted–Evans–Polanyi principle emerged as a fundamental framework, revealing that activation energies of elemental reactions on catalyst surfaces are proportional to adsorption energies [27]. DFT calculations have demonstrated remarkable precision, achieving adsorption energy predictions within 0.1 eV [28], and successfully applying this strategy to HER and oxygen reduction reactions with exceptional agreement between theoretical and experimental results [24,29].

Despite these significant advancements, substantial challenges persist. Existing HER alloy catalyst datasets are often acquired at high computational costs, and current feature extraction methods may not comprehensively capture the nuanced catalytic properties. These limitations can significantly compromise model accuracy and restrict the exploration of potential catalytic materials.

Our research specifically aims to develop novel machine learning methodologies for predicting and designing high-performance, cost-effective intermetallic catalysts for HER and OER. By integrating advanced computational techniques with fundamental electrochemical principles, we seek to create a systematic, data-driven approach to catalyst design that can overcome the limitations of traditional trial-and-error methodologies. The path to a sustainable hydrogen economy is complex and multifaceted, requiring innovative solutions that transcend disciplinary boundaries. Through advanced machine learning approaches, we stand at the cusp of a potential breakthrough that could revolutionize green hydrogen production, bringing us significantly closer to a decarbonized, sustainable global energy system.

**Results and discussions**

**Dataset Composition and Characteristics**

Our research uses a comprehensive dataset extracted from the Catalysis-hub database, representing a significant compilation of alloy catalysis data for hydrogen evolution reaction (HER) and oxygen evolution reaction (OER). The dataset encompasses a total of 16,226 distinct data points, with 8,856 entries specifically focused on HER catalysis and 7,370 entries dedicated to OER catalysis[30].
The dataset provides a multifaceted representation of alloy catalytic properties, incorporating critical parameters such as Gibbs free energy of adsorption ($\Delta G$), detailed chemical compositions, and specific adsorption sites.
A distinctive feature of the dataset is its extensive coverage of chemisorption properties across nine critical adsorbates, including Carbon (C), Oxygen (O), Nitrogen (N), Hydrogen (H), Sulfur (S), Carbonaceous species (CH x), Hydroxyl (OH), Amine (NH), and Sulfhydryl (SH) groups. These adsorbates were systematically analyzed across 2,035 unique bimetallic alloy surfaces, exploring five distinct stoichiometric ratios: 0%, 25%, 50%, 75%, and 100%[30]. The structural complexity of the dataset is further enhanced by its representation of diverse alloy configurations. Specifically, the data incorporates $L1_0$ and $L1_2$ Strukturbrecht designations, corresponding to face-centered cubic (FCC) crystal structures with AB and $A_3B$ stoichiometries, respectively.

The dataset encompasses 37 elemental compositions, providing a broad spectrum of alloy materials for both HER and OER analyses. This extensive compositional range enables a robust exploration of structure-property relationships in electrocatalytic systems. Notably, the dataset reveals that different adsorption sites can exhibit markedly distinct capabilities for HER and OER, highlighting the complex interplay between surface geometry, chemical composition, and catalytic activity.

**Composition and site based features**

The feature generation process was a critical component of our computational catalyst design methodology, aimed at transforming raw material data into meaningful predictive features. We used a comprehensive approach to feature extraction, using advanced computational tools and statistical techniques to capture the complex characteristics of potential electrocatalysts. Compositional features were systematically extracted using the Matminer package, a specialized Python library for materials informatics[31]. Matminer enabled the transformation of raw chemical composition data into a rich set of numerical descriptors that contains the fundamental electronic, structural, and compositional properties of the catalysts. This approach goes beyond traditional feature extraction, providing a detailed representation of material characteristics that are crucial for predicting catalytic performance.

The site-related features underwent a sophisticated preprocessing pipeline. Initially generated through one-hot encoding, these features initially comprised 20 distinct descriptors capturing site-specific attributes. Recognizing the challenge of high-dimensional data, we applied Principal Component Analysis (PCA) to reduce dimensionality while preserving critical information. The PCA transformation effectively reduced the 20 original site features into a more manageable set of 5 principal components, significantly mitigating potential overfitting and computational complexity.

To ensure the independence and reliability of our feature set, we implemented Pearson correlation analysis. This method identified and removed highly correlated features, further refining our feature space and minimizing potential multicollinearity issues. The correlation analysis is crucial in machine learning applications, as it helps create a more robust and interpretable feature representation. The final feature set represented a carefully selected collection of material descriptors, totaling 80 features for the Hydrogen Evolution Reaction (HER) and 76 features for the Oxygen Evolution Reaction (OER). This feature engineering approach provides a solid foundation for our machine learning models based on compositional, structural, and site-specific characteristics that govern catalytic performance. The deatiled workflow digram of the ML methodology is shown in figure 1.

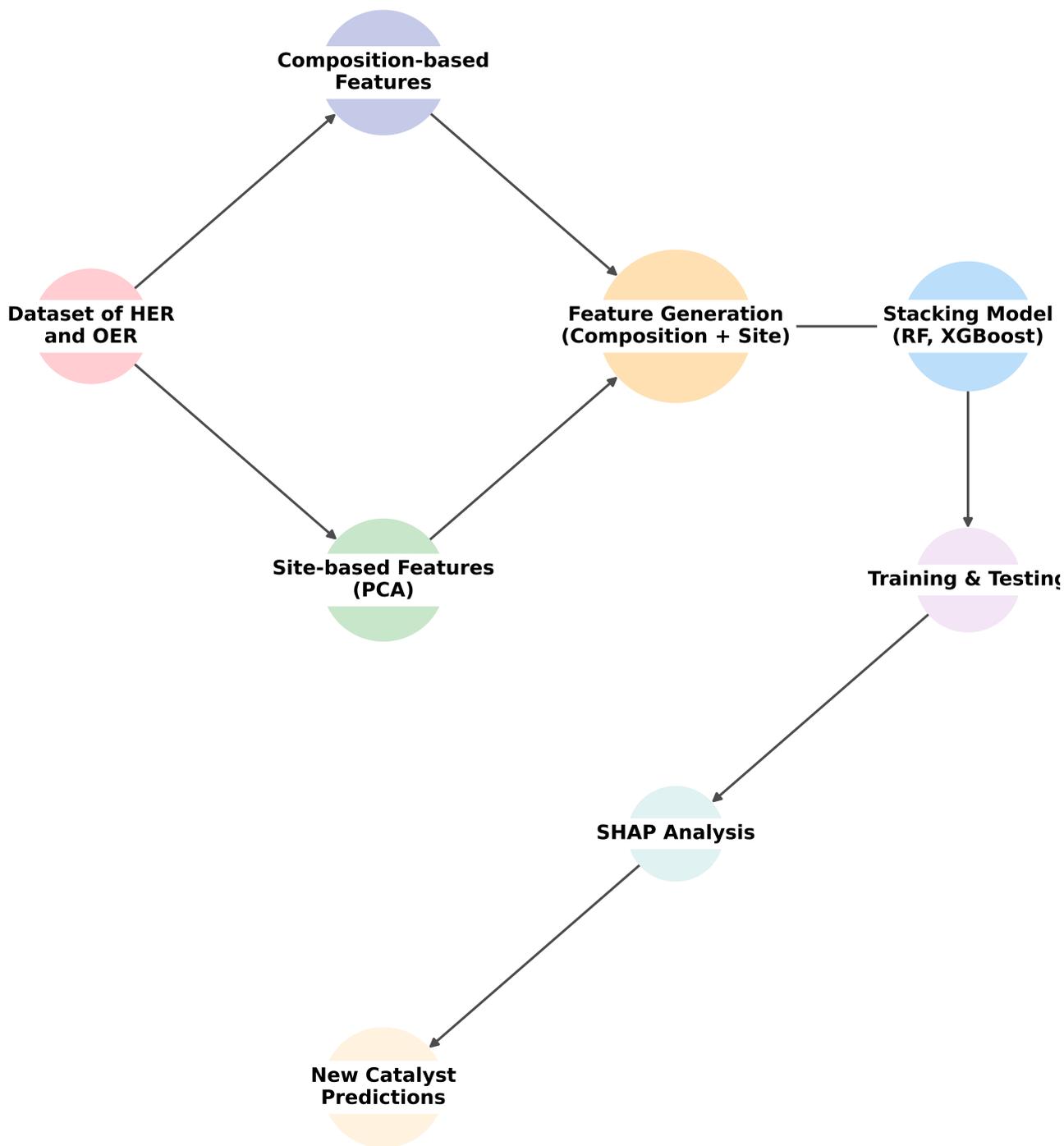

*Figure 1: Showing ML methodology adopted for Hydrogen Evolution Reaction (HER) and Oxygen Evolution Reaction (OER)*

**ML model**

For the computational approach to catalyst performance prediction, we used a sophisticated ensemble learning technique known as stacking, a advanced meta-learning methodology that transcends traditional predictive modeling strategies[32]. By integrating multiple machine learning algorithms, stacking offers a new approach to capturing the complex relationships in catalytic material properties.

Our stacking model was constructed using Random Forest (RF)[33] and XGBoost[34] as base models, strategically selected for their complementary predictive capabilities. Random Forest, with its ensemble of decision trees, provides robust handling of non-linear relationships, while XGBoost brings advanced gradient boosting techniques that excel in capturing complex data interactions. These base models were designed to leverage their individual strengths, creating a comprehensive approach to feature interpretation. Support Vector Regression (SVR)[35] was employed as the meta-model, serving as a high-level learner that synthesizes predictions from the base models. This approach allows for a complex integration of diverse predictive signals, enabling the model to understand deeper insights from the catalytic dataset. The meta-model effectively learns the optimal method of combining base model predictions, uncovering subtle relationships that individual models might fail to detect.

The dataset underwent a rigorous partitioning process, with 90% allocated to training and 10% reserved for independent testing. This division ensures robust model validation while mitigating the risk of overfitting. Performance evaluation focused on two critical metrics: R-squared ($R^2$), which quantifies the proportion of variance explained by the model, and Mean Absolute Error (MAE), providing a direct measure of prediction accuracy. $R^2$ and MAE values of 0.98 and 0.251 was achieved for HER and $R^2$ and MAE values of 0.94 and 0.121 was achieved for OER.

Visualization of the model's predictive capabilities was realized through comprehensive training and testing diagrams for both Hydrogen Evolution Reaction (HER) and Oxygen Evolution Reaction (OER), presented in Figure 2.

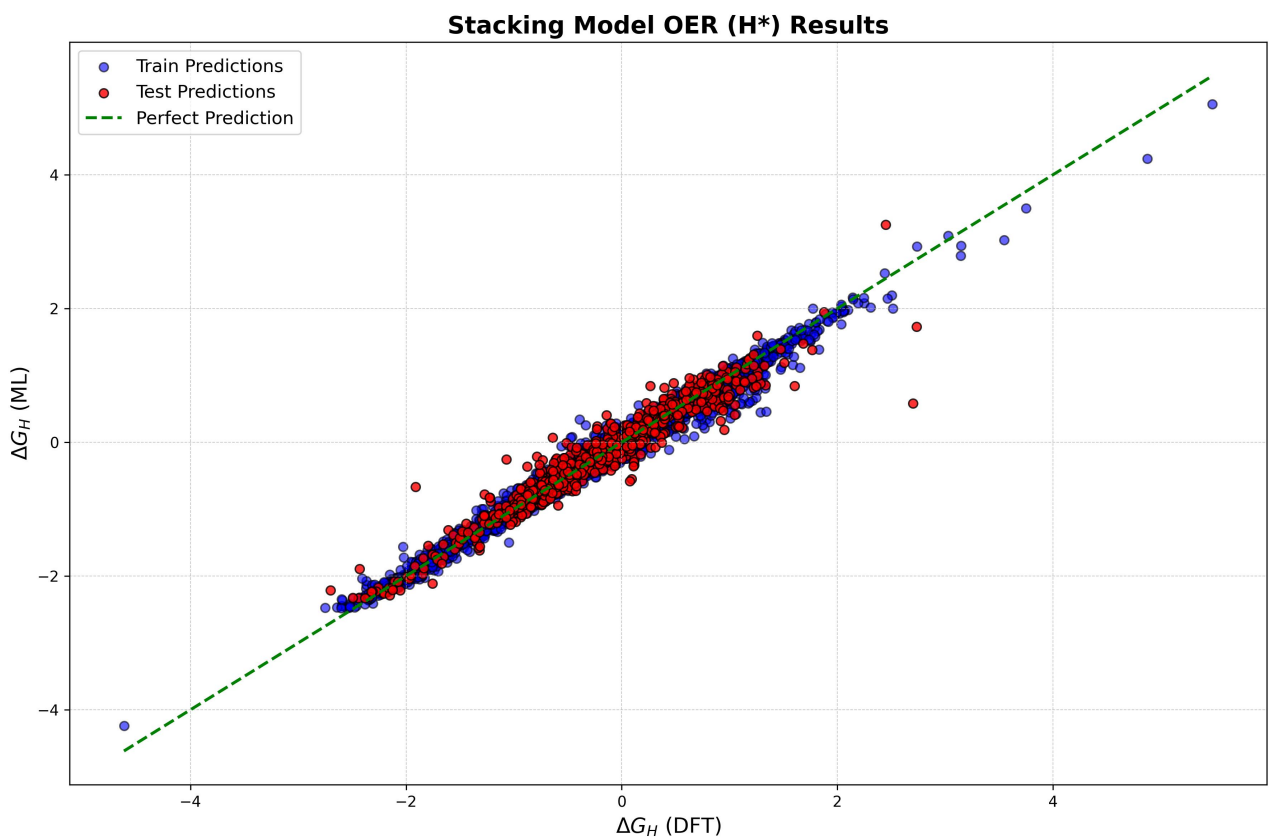

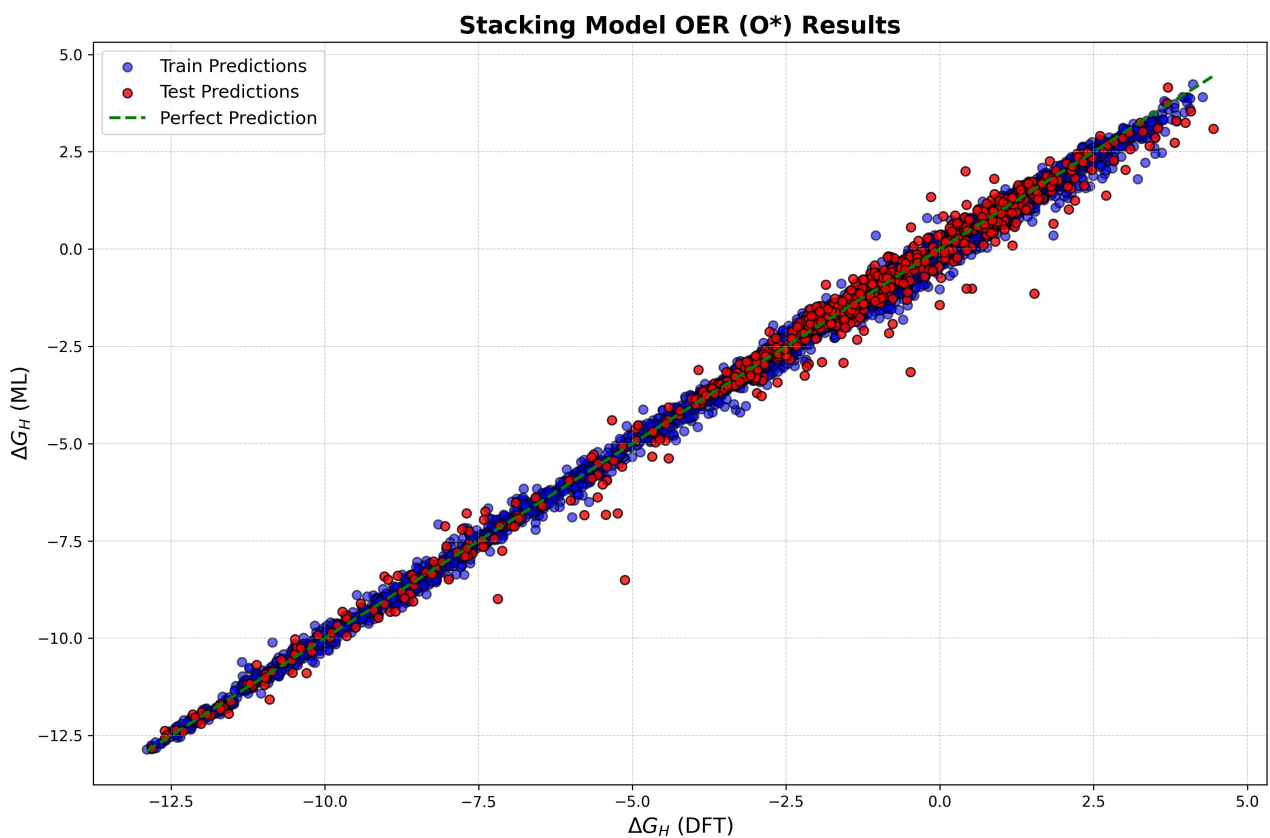

*Figure2: Training and testing predictions with the Stacking model for both Hydrogen Evolution Reaction (HER) and Oxygen Evolution Reaction (OER)*

**Feature importance for HER**

The study of feature importance in predicting catalytic performance for the Hydrogen Evolution Reaction (HER) provides crucial insights into the interplay between various material descriptors and Gibbs free energy (ΔG) of adsorption. Shapley Additive Explanations (SHAP)[36] analysis was employed to systematically identify and interpret the most influential features driving the model's predictions. The results of this analysis, depicted in the SHAP summary plot, reveal the critical contributions of several compositional and structural features.

Among the most significant predictors, the "MagpieData mean MeltingT" stands out as a dominant feature, reflecting the average melting temperature of alloy components. This parameter's relevance suggests that alloys with higher melting temperatures exhibit favorable stability and robust structural properties, which are crucial for HER performance. Additionally, "PCA_Site_3," a principal component derived from site-specific descriptors, emerges as a significant feature. This indicates the importance of adsorption site attributes, emphasizing the structural and geometric factors influencing the interaction between catalytic surfaces and hydrogen adsorbates.

The "MagpieData mean MendeleevNumber" also demonstrates high SHAP values, signifying the role of periodic table trends, such as electronegativity and atomic size, in determining alloy catalytic behavior. Notably, features like "MagpieData mode NdUnfilled" and "MagpieData mode NUnfilled," which capture electronic configuration attributes such as the number of unfilled electronic states, underscore the influence of electronic structure on catalytic activity. This aligns with the understanding that electronic properties govern adsorption energetics and catalytic surface interactions.

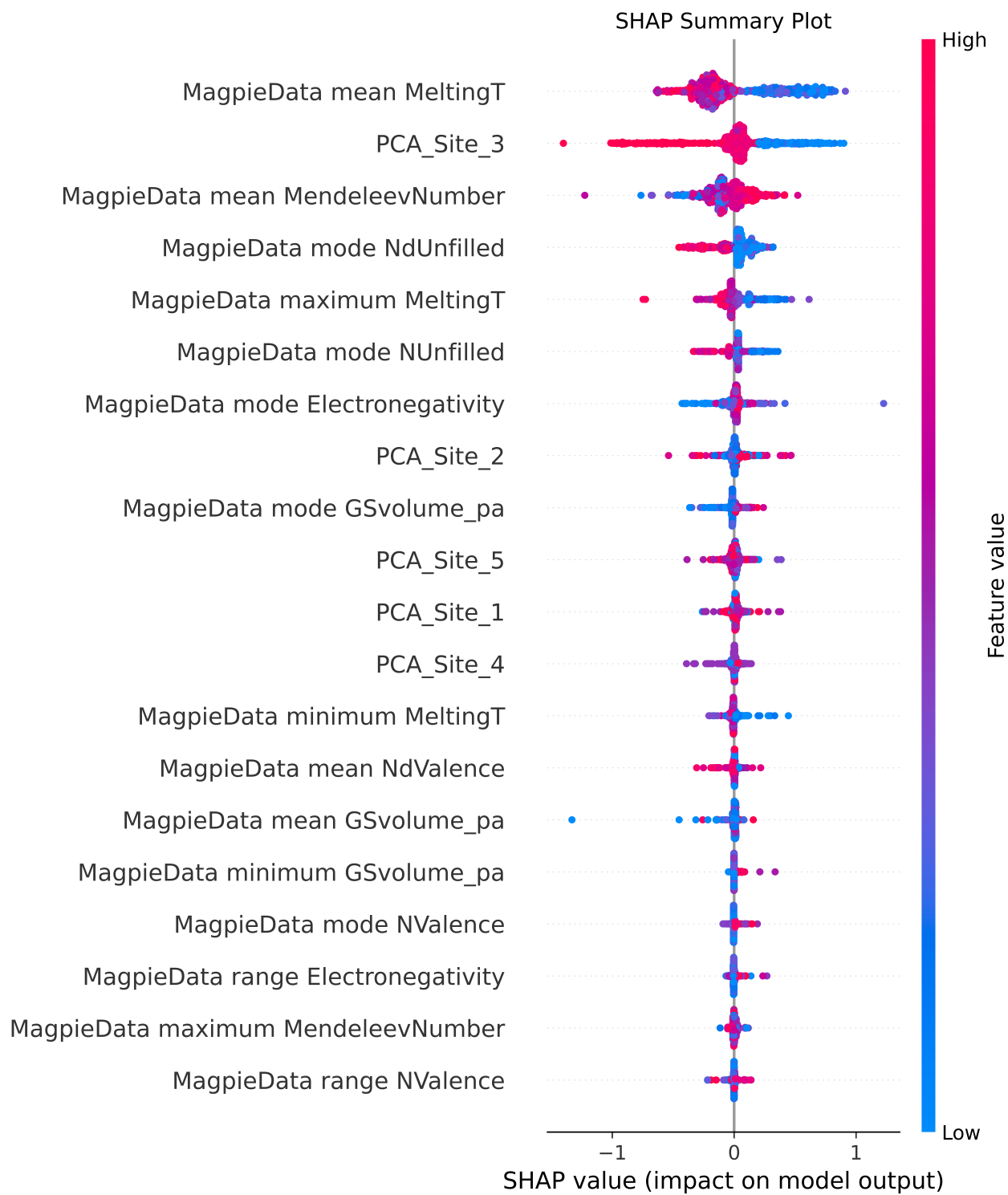

Figure 3: The SHAP summary plot for Hydrogen Evolution Reaction (HER)

The maximum and minimum values of melting temperatures further highlight the interplay of compositional extremes in alloys. These features suggest that the variance in melting points within an alloy influences its Gibbs free energy landscape, dictating the strength and stability of hydrogen adsorption. Similarly, "MagpieData mode Electronegativity" and "MagpieData range

Electronegativity" emphasize the role of electronic uniformity and variation within alloy compositions in tuning catalytic efficiency.

Principal components representing site-specific features, such as "PCA_Site_2" and "PCA_Site_5," capture critical geometric and spatial characteristics, further reinforcing the significance of adsorption site morphology. The inclusion of site features reduced through PCA not only mitigates dimensionality but also ensures that essential structural details are preserved, helping in robust prediction of Gibbs free energy variations.

These insights collectively underscore the complex and multifactorial nature of HER catalysis, where compositional, structural, and electronic descriptors synergistically determine the catalytic performance. By linking these features to the fundamental thermodynamic parameter ($\Delta G$) of hydrogen adsorption, the SHAP analysis provides a comprehensive understanding of the critical material properties required for designing high-performing HER catalysts.

**Feature importance for OER**

The SHAP analysis reveals key features influencing the predictive model for Oxygen Evolution Reaction (OER) catalytic activity. Among the most significant contributors are the principal component analysis (PCA) features, specifically PCA_Site_3 and PCA_Site_2, which demonstrate substantial impact on the model output. These features likely encapsulate the geometric as well as electronic attributes of the active catalytic sites, underscoring their importance in modulating the adsorption and activation of oxygen intermediates during the OER process. In addition to PCA-derived features, atomic and electronic properties also exhibit notable influence. The MagpieData maximum CovalentRadius emerges as a key descriptor, suggesting that larger covalent radii within the alloy composition enhance catalytic performance. This relationship could be attributed to increased bond flexibility and optimized interaction between oxygen intermediates and the catalytic surface. Similarly, MagpieData mean NdValence and MagpieData mode Column reflect the significance of electronic configurations and elemental group tendencies in determining catalytic behavior.

Other descriptors, such as MagpieData minimum NdValence and MagpieData minimum MendeleevNumber, further highlight the interplay between electronic properties and elemental identity in fine-tuning catalytic activity. These observations indicate that both intrinsic material properties and site-specific geometric features play complementary roles in governing OER efficiency.

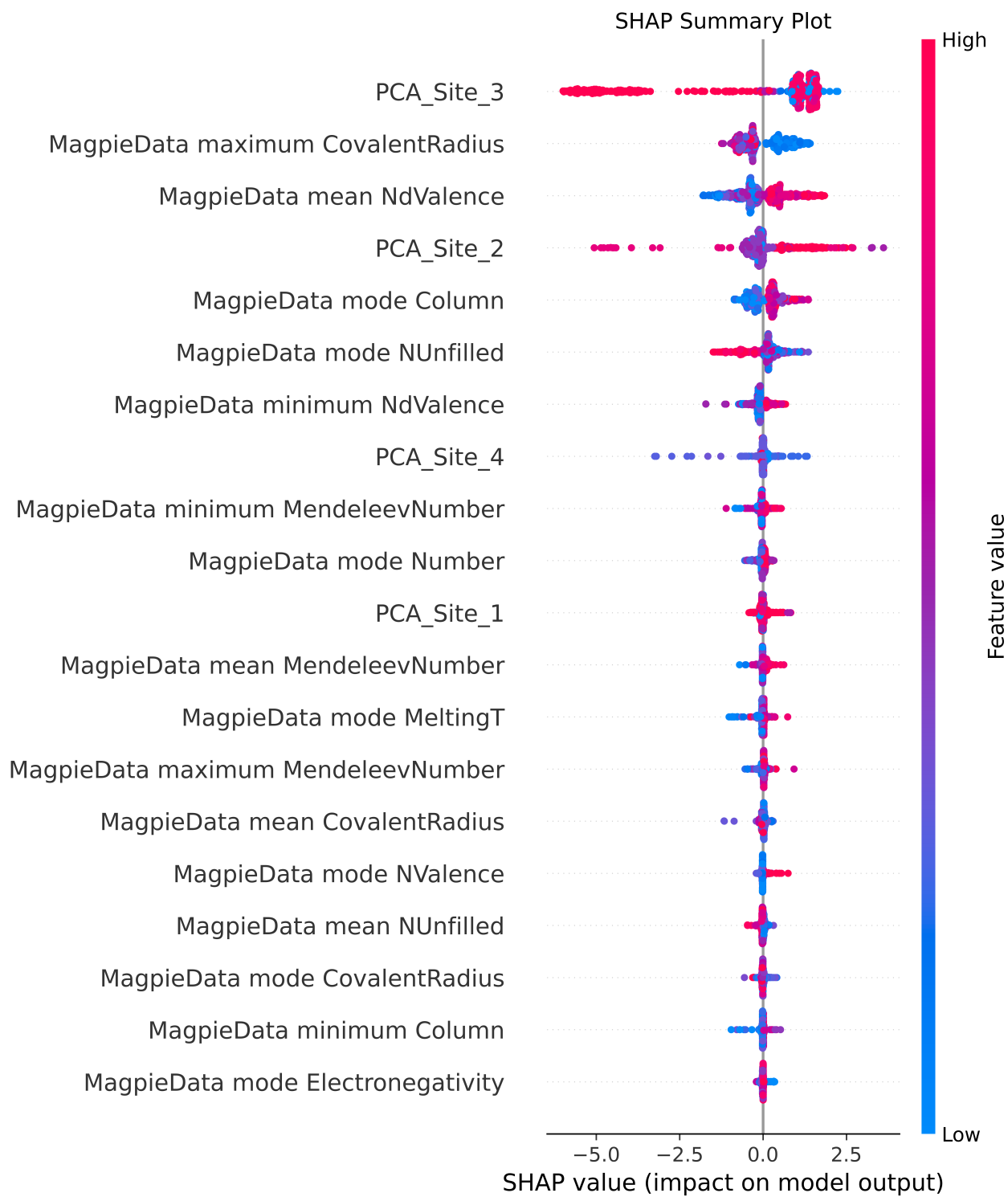

*Figure 4: The SHAP summary plot for Oxygen Evolution Reaction (OER)*

**Conclusion**

The quest for sustainable energy solutions demands innovative approaches that can fundamentally transform our understanding of catalytic materials. Our research represents a significant stride towards this goal, demonstrating the immense potential of data-driven methodologies in electrocatalyst design. By integrating advanced machine learning techniques with feature engineering, we have developed a computational framework that offers unprecedented insights into the complex catalytic performance. The stacking ensemble model developed in this study goes beyond traditional predictive approaches, capturing the relationships between material properties and catalytic activity. Our comprehensive analysis showed that catalytic performance emerges from a delicate interplay of compositional, structural, and electronic characteristics. Features such as melting temperature, electronic configuration, and adsorption site geometry collectively determine the efficiency of hydrogen and oxygen evolution reactions. The exceptional predictive performance of our model—with $R^2$ values approaching 0.98 for HER and 0.94 for OER—underscores the power of machine learning in unraveling complex materials science challenges. The Shapley Additive Explanations (SHAP) analysis provided critical insights into the most influential descriptors, highlighting the importance of both intrinsic material properties and site-specific geometric features. Perhaps most significantly, this research offers a generalizable framework for computational catalyst design. By demonstrating the ability to predict catalytic performance with remarkable accuracy, we open new pathways for accelerating materials discovery. The approach presented here can be extended to explore a wide range of electrocatalytic systems, potentially revolutionizing our approach to sustainable energy technologies. As the global community confronts the urgent challenge of decarbonization, computational approaches like those presented in this study become increasingly critical. By bridging advanced computational techniques with fundamental electrochemical principles, we move closer to developing the next generation of high-performance, cost-effective catalysts that can transform our energy landscape. The journey towards sustainable hydrogen production is complex and multifaceted. Our research provides a powerful computational compass, guiding researchers towards more efficient, innovative catalyst designs. It represents not just a scientific advancement, but a crucial step towards a more sustainable, decarbonized global energy future.

AUTHOR INFORMATION:

Corresponding Author: ph18d200@smail.iitm.ac.in

CONFLICT OF INTEREST

The authors declare no competing interest.

DATA AVAILABILITY

The data that support the findings of this study are available from the corresponding authors upon reasonable request.

**Acknowledgements:**

We would like to acknowledge the use of the computing resources at High Performance Computing Environment, Indian Institute of Technology Madras.